\documentclass{aa}
\usepackage{epsfig}
\usepackage{psfig}
\usepackage{graphicx}
\usepackage{txfonts}
\usepackage{lscape}

\titlerunning{X-ray observation of PSR~B1957$+$20}

\begin{document}

  \title{XMM-Newton Observations of the Black Widow Pulsar PSR B1957+20}
  \author{Hsiu-Hui Huang \and Werner Becker}
  \institute{Max-Planck-Institut f\"{u}r extraterrestrische Physik,
              Giessenbachstrasse, 85748 Garching, Germany}

  \date{Received October 13, 2006; accepted January 9, 2007}

  \abstract{We report on XMM-Newton observations of the ``Black Widow pulsar", PSR B1957+20.
   The pulsar's X-ray emission is non-thermal and best modeled with a single powerlaw spectrum   
   of photon index $2.03^{+0.51}_{-0.36}$. No coherent X-ray pulsations at the pulsar's 
   spin-period could be detected, though a strong binary-phase dependence of the X-ray flux 
   is observed for the first time. The data suggest that the majority of the pulsar's X-radiation 
   is emitted from a small part of the binary orbit only. We identified this part as being near 
   to where the radio eclipse takes place. This could mean that the X-rays from PSR B1957+20 are 
   mostly due to intra-shock emission which is strongest when the pulsar wind interacts with the 
   ablated material from the companion star.

   }

   \maketitle


\section{Introduction}

Until now, more than 1700 rotation-powered radio pulsars are detected. Among 
them are about 10\% which are millisecond pulsars (MSPs) (Manchester et al.~2005).
They form a separate population. The majority of them resides in Globular 
Clusters (c.f.~Bogdanov et al.~2006). MSPs are presumed to have been spun up 
in a past accretion phase by mass and angular transfer from a binary companion 
(Alpar et al.~1982).  Only about one third of them are seen to be solitary. It 
is believed that they lost their companion, e.g.~in a violent supernova event. 
All MSPs possess very short spin periods of less than 20 ms and show a high 
spin stability with period derivatives in the range $\approx 10^{-18}-10^{-21}$. 
MSPs are generally very old neutron stars with spin-down ages $\tau = P/2\dot{P}$ 
of $\sim 10^{9}-10^{10}$ years and low surface magnetic fields in the range 
$B ~\propto  \sqrt{(P \dot{P})} \sim 10^{8} - 10^{10}$ G.  

At present, about 50\% of all X-ray detected rotation-powered radio pulsars are MSPs
(c.f.~Bogdanov et al.~2006 and references therein). Among them an extraordinarily 
rich astrophysics binary system which is formed by the millisecond pulsar 
PSR B1957+20 and its 0.025 $M_{\odot}$ low mass white dwarf companion (Fruchter, 
Stinebring, $\&$ Taylor 1988). The binary period of the system is 9.16 hours.
The spin period of the pulsar is 1.6 ms which is the third shortest among all known MSPs. 
Its period derivative of $\dot{P} = 1.69 
\times 10^{-20} ~s~s^{-1}$ implies a spin-down energy of $\dot{E} = 10^{35} 
~erg~s^{-1}$, a characteristic spin-down age of $ > 2 \times 10^{9}$ years, and 
a dipole surface magnetic field of $B_\perp = 1.4 \times 10^{8}$ Gauss. 
Optical observations 
by Fruchter et al.~(1988) and van Paradijs et al.~(1988) revealed that the pulsar 
wind consisting of electromagnetic radiation and high-energy particles is ablating 
and evaporating its white dwarf companion star. This rarely observed property 
gave the pulsar the name {\em black widow pulsar}.  Interestingly, the radio 
emission from the pulsar is eclipsed for approximately 10\% of each orbit by 
material expelled from the white dwarf companion. For a radio dispersion 
measure inferred distance of 1.5 kpc (Taylor $\&$ Cordes 1993) the pulsar 
moves through the sky with a supersonic velocity of 220 km/sec. The interaction 
of a relativistic wind flowing away from the pulsar with the interstellar 
medium (ISM) produces an H$_\alpha$ bow shock which was the first one seen 
around a ``recycled" pulsar (Kulkarni \& Hester 1988).

In 1992 Kulkarni et al.~published a contour map of the X-ray emission of 
PSR B1957+20 which was derived from ROSAT PSPC observations. Although this 
ROSAT data were very sparse in statistics it led the authors to predict 
faint diffuse X-ray emission with constant surface brightness to be present 
along a cylindrical trail formed when the relativistic pulsar wind expands 
into pressure equilibrium with the interstellar medium behind the nebula. 
The much improved sensitivity of the Chandra and XMM-Newton observatories 
made it possible to probe and investigate the structure and properties of 
this unique binary system in much higher detail than it was possible with 
ROSAT, ASCA or BeppoSAX. A narrow X-ray tail with the extent of 16 arcsec 
and the orientation to the north-east was detected from it in deep Chandra 
observations by Stappers et al.~(2003). 
 
Searching in ROSAT data for a modulation of the pulsar's X-ray emission 
as a function of its orbital phase revealed a suggestive but insignificant 
increases in flux before (at phase $\phi \sim 0.17$) and after (at phase 
$\phi \sim 0.4 - 0.5$) the pulsar radio eclipse ($\phi = 0.25$) (Kulkarni 
et al.~1992). Taking Chandra data into account revealed a hint that the 
lowest and highest fluxes are located during and immediately after the 
radio eclipse, respectively. The statistical significance of this 
modulation observed by Chandra, though, is only 98\% and thus prevents 
any final conclusion on it (Stappers et al.~2003).  

In this paper we report on XMM-Newton observations of PSR B1957+20 and
its white dwarf companion. The paper is organized as follows. In \S2 we 
describe the observations and data analysis. \S3 gives a summary and 
discussion.


\section{Observations and Data Analysis}

PSR B1957+20 was observed with XMM-Newton on October 31, 2004 for a 
30 ksec effective exposure. In this observation, the EPIC-MOS1 and 
MOS2 instruments were operated in full-frame mode using the thin 
filter to block optical stray light. The EPIC-PN detector was setup 
to operate in the fast timing mode. Because of the reduced spatial 
information provided by the PN in timing mode we use the MOS1/2 data 
for imaging and spectral analysis of the pulsar and its diffuse 
X-ray nebula while the PN data having a temporal resolution of 
0.02956 ms allowed us to search for X-ray pulsations from the 
pulsar. All the data were processed with the XMM-Newton Science 
Analysis Software (SAS) package (Version 6.5.0). Spatial and spectral
analyses were restricted to the $0.3-10.0$ keV energy band while for timing 
analysis events were selected for the energy range $0.3-3.0$ keV.

\subsection{Spatial Analysis}
Figure 1 shows the combined EPIC-MOS1/2 image of the PSR B1957+20 system.
The image was created with a binning factor of 6 arcsec and by using an
adaptive smoothing algorithm with a Gaussian kernel of $\sigma < 4$ pixels
in order to better make visible faint diffuse emission. The extent of the 
diffuse emission with its orientation to the north-east is about 16 arcsec 
which is consistent with the previous result derived from the Chandra 
Observations. However, the detailed structure of the X-ray emission from 
XMM-Newton can not be as clearly seen as from Chandra due to the 10 times 
wider Point Spread Function (PSF) of XMM-Newton. 

Inspecting the XMM-Newton MOS1/2 image two faint features (denoted as A and B) 
which contribute to only about 3\% of the total X-ray flux are apparent. In order 
to investigate whether this faint features are associated with nearby stars we 
inspected the Digitized Sky Survey data (DSS) and the USNO-B1.0 Catalogue for 
possible sources. These catalogues which are limited down to 22 mag (Krongold, 
Dultzin-Hacyan \& Marziani 2001) and 21 mag (Monet et al.~2003) do not reveal 
possible counterparts. These features are not seen in the Chandra image though 
(Stappers et al.~2003).

   \begin{figure}
   \centerline{\psfig{figure=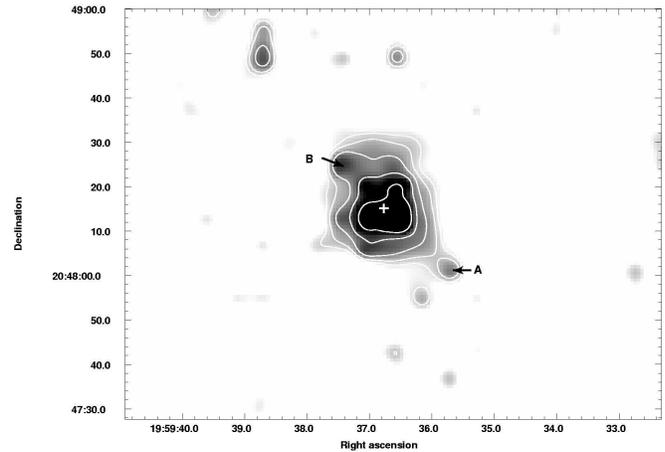,width=9cm}}
   \caption{XMM-Newton MOS1/2 image of the PSR B1957+20 -WD system with contour lines overlaid. 
   The contour lines are at the levels of  $(5.1, 6.1, 9.4, 16.3, 32.4, 69.0) \times 10^{-6}\,
   \mbox{cts s}^{-1}\mbox{arcsec}^{-2}$. The position of the pulsar is indicated.}\label{Fig1}
   \end{figure}

\subsection{Spectral Analysis}
Combined EPIC-MOS1/2 data of PSR B1957+20 were extracted from a circle of 30 arcsec
radius centered at the pulsar position (RA (J2000) $= 19^{h}59^{m}36^{s}.77$, 
Dec $= 20^{\circ}48'15".12$). The selection region contains about 85\% of all source 
counts. Background photons were selected from a source-free region near to the pulsar 
position. Response files were derived by using the XMM-Newton SAS tasks RMFGEN and ARFGEN. 

After subtracting background photons, in total 338 sources counts were available for a 
spectral analysis. The extracted spectra were binned with at least 30 source counts 
per bin. Assuming that the emission originates from the interaction of the pulsar 
wind with the ISM or with the stellar wind we expect synchrotron radiation to be the 
emission mechanism of the detected X-rays. To test this hypothesis we fitted the 
spectrum with a power law model. Indeed, this model describes the observed spectrum 
with a reduced $\chi^{2}_{\nu}$ of 1.09 (for 8 D.O.F.). The photon-index is found 
to be $\alpha = 2.03^{+0.51}_{-0.36}$. The column absorption $N_{H}$ is $8.0 \times 
10^{20} ~cm^{-2}$.  For the normalization at 1 keV we find $1.5^{+0.9}_{-0.3} 
\times 10^{-5}\,\mbox{photons keV}^{-1} \mbox{cm}^{-2} \mbox{sec}^{-1}$ (1-$\sigma$ 
confidence for 1 parameter of interest). The spectrum (data and model) and the fit 
residuals are shown in Figure 2. 

The unabsorbed X-ray fluxed derived from the best fitting model parameters is 
$f_{x}=8.35 \times 10^{-14} ~erg ~s^{-1}~cm^{-2}$ and $f_{x}=7.87 \times 
10^{-14} ~erg ~s^{-1}~cm^{-2}$ in the $0.3-10$ keV and $0.1-2.4$ keV 
energy band, respectively. The X-ray luminosities in these energy bands -- calculated 
for a pulsar distance of 1.5 kpc -- are $L_{x}(0.3 - 10.0\mbox{keV}) 
= 2.24 \times 10^{31}~erg~s^{-1}$ and $L_{x}(0.1 - 2.4 \mbox{keV}) 
= 2.12 \times 10^{31}~erg~s^{-1}$, respectively. The conversion 
efficiency $L_{x}/\dot{E}$ in the $0.1-2.4$ keV band is found to be $\sim 2.12 
\times 10^{-4}$. 

In order to check whether the spectral emission characteristics changes for
photons detected in a smaller compact region of 10 arcsec radius at the pulsar 
position we applied a spectral fit to this events only. The encircled
energy within 10 arcsec is 60 $\%$. In total,  186 sources counts were available
for spectral fits. We did not find any significant change in the spectral 
parameters than reported above. 

   \begin{figure}
   \centerline{\psfig{figure=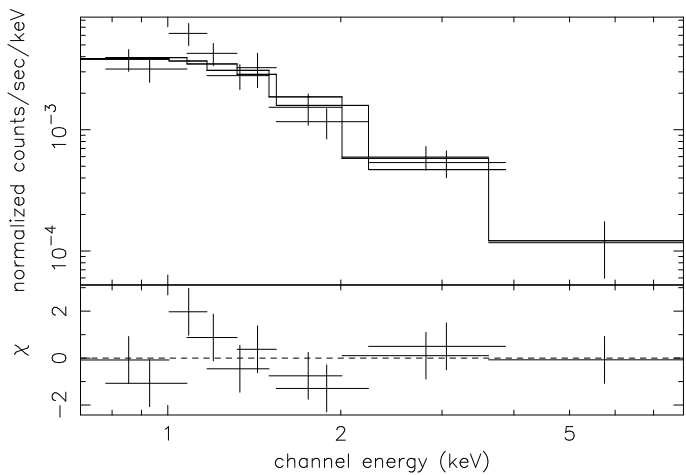,angle=-90,width=9cm}}
   \caption{Energy spectrum of PSR B1957+20 obtained from the XMM-Newton MOS1/2 data. 
    The plot shows the X-ray spectrum fitted with an absorbed power law model 
    (upper panel) and contribution to the $\chi^2$ fit statistic (lower panel).}
   \end{figure}

\subsection{Timing Analysis}

 The EPIC-PN camera observed the pulsar in the fast timing mode. In this mode 
 the spatial and spectral information from a $64 \times 199$ CCD pixel 
 array is condensed into a one dimensional $64 \times 1$ pixel array (1D-image), 
 i.e.~the spatial information in Y-direction is lost due to the continuous 
 read-out of the CCD. The complete photon flux (source plus DC emission from
 foreground or background sources located along the  read-out direction) is 
 accumulated and collapsed in the final 1D-image, severely reducing the 
 signal-to-noise ratio of pulsed emission and preventing the detection of 
 weak X-ray pulsations from the target of interest.

 In order to search for X-ray pulsations from PSR B1957+20 we extracted
 1639 counts from the CCD columns $33- 41$ in which the pulsar got located. 
 In order to increase the signal-to-noise ratio we restricted the analysis 
 to the energy range $0.3-3.0$ keV. Below and beyond this energy 
 band the accumulative sky and instrument background noise exceeds the contribution 
 from the pulsar itself (c.f.~Becker \& Aschenbach ~2002). However, still 80 $\%$ of
 the counts are estimated to be derived from the background. 

 The photon arrival times were corrected to the solar system barycentre 
 with the BARYCEN tool (version: 1.17.3, JPL DE200 Earth ephemeris) of the 
 SAS package. As the pulsar is in a binary we corrected for the orbital 
 motion of the pulsar by using the method of Blandford \& Teukolsky (1976). 

 As millisecond pulsars are known to be extremely stable clocks we used the 
 pulsar ephemeris from the ATNF Catalogue, $f = 622.122030511927$ ~Hz and 
 $\dot{f} = -6.5221 \times 10^{-15}\mbox{sec}^{-2}$ (at MJD = 48196.0) to 
 perform a period folding. By using the $Z^{2}_{n}$ statistics (Buccheri 
 et al.~1983) with the harmonics number (n) from one to ten no significant 
 signal was detected at the radio spin period extrapolated for the epoch of
 the XMM-Newton observation. Restricting the period search to the various
 smaller energy bands did not change the result. A pulsed fraction upper 
 limit of 9\% (1-$\sigma$) is deduced by assuming a sinusoidal pulsed profile.

 Arons $\&$ Tavani (1993) predicted that depending on the flow speed and the 
 degree of absorption and/or scattering by the companion wind the X-ray
 emission from PSR B1957+20 increases by up to a factor of 2.2 at the 
 orbital phase  before and after the radio eclipse. In order to test this
 prediction we created a lightcurve by binning all events in bins of 
 1.5 ksec width. With an effective exposure time of about 30 ksec and 
 an orbit period of 9.16-h the XMM-Newton MOS1/2 and PN data cover roughly 
 83 $\%$ and 92 $\%$ of one binary orbit, respectively.
 Table 1 lists the first and the last photon arrival times recorded by 
 the MOS1/2 and EPIC-PN detectors and the corresponding pulsar binary orbital phase.
 The lightcurves resulting from the EPIC-PN and MOS1/2 data are 
 shown in Figure 3. For the lightcurve deduced from the PN-data (upper 
 solid curve) it is clearly seen that before the radio eclipse, 
 i.e.~between the orbital phase 0.1 and 0.25, the X-ray emission 
 increases. The emission in the highest bin is about a factor of 3.0
 stronger than at other orbital phase angles before and after the radio 
 eclipse. This is in agreement with the predictions made by Arons $\&$ 
 Tavani (1993). 

 Although it is the first time that a significant X-ray flux modulation 
 from PSR B1957+20 is observed, the flux increase near the phase of the 
 radio eclipse is virtually only seen in the PN data. Owing to the short 
 observation time of 30 ksec, which is less than the time of one full 
 binary orbit, the orbit angles $0.30 - 0.38$ and $0.18 - 0.36$ are not 
 covered by EPIC-PN and MOS1/2 data at all. 
 It thus turns out that it was a great fortune that the
 EPIC-PN, more or less by chance, covered the orbital phase range of the 
 radio eclipse and provided us evidence for the flux enhancement while 
 the MOS1/2 detectors were already switched off. 

   \begin{figure}
   \centerline{\psfig{figure=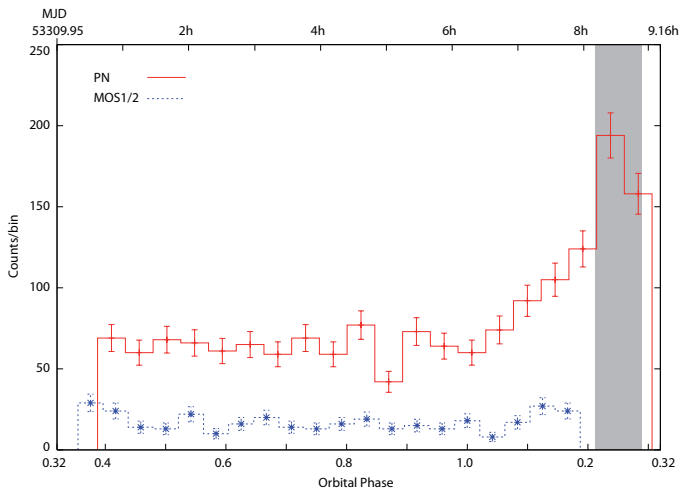,width=9cm}}
   \caption{X-ray emission from PSR B1957+20 within $0.3-3.0$ keV as function of 
    the pulsar's orbital phase ($\phi$). We mapped one complete orbital 
    period of this system at the starting point of $MJD=53309.95$, 
    i.e. $\phi = 0.32$. $\phi = 1.0$ corresponds to the ascending node 
    of the pulsar orbit. The upper curve was obtained from 
    the XMM-Newton EPIC-PN (background level at 66 cts/bin). The lower 
    lightcurve is obtained from the MOS1/2 data. 
    The gray strip between the orbital angle $0.21 - 0.29$ indicates the eclipse
    of the pulsar. Phase bins with zero
    counts correspond to phase angles not covered in the observation.} 
   \end{figure}

\begin{table*}
\caption{List of the first and the last photon arrival times and its corresponding 
orbital phase of PSR B1957+20 for the MOS1/2 and PN detectors. Arrival times are 
corrected to the solar system barycentre. The orbital phase is measured from the 
time of ascending node.} \label{table:1}
\centering                          
\begin{tabular}{c c c c c c}        
\hline\hline                 
 Date set &\multicolumn{2}{c}{1st photon}&\multicolumn{2}{c}{last photon} & Duration (s)\\
\cline{2-5}
 &Time (MJD) & Orbital Phase & Time (MJD) & Orbital Phase \\
\hline                        
   MOS1 & 53309.9688 & 0.3605 & 53310.2830 & 0.1830 & 27143.4\\
   MOS2 & 53309.9666 & 0.3548 & 53310.2846 & 0.1871 & 27467.1\\
   PN & 53309.9791 & 0.3873 & 53310.3301 & 0.3063 & 30327.4 \\
\hline                                
\end{tabular}
\end{table*}


\section{Summary and Discussion}

Interaction between relativistic pulsar winds which carry away the 
rotational energy of pulsars and the surrounding medium is expected 
to create detectable X-ray emission. Indeed, there are about 30 
pulsar wind nebulae (PWNe) currently detected in the X-ray band 
(e.g.~Kaspi, Roberts \& Harding 2006, Gaensler $\&$ Slane 2006, 
Kargaltsev $\&$ Pavlov 2006). However, these PWNe are all powered by young 
and powerful pulsars with spin-down energies of more than 
$\sim 3.6 \times 10^{36}~erg~s^{-1}$.
Until now, only two MSPs are known to have X-ray nebulae. 
They are PSR B1957+20 (Stappers et al.~2003) and PSR J2124-3358 
(Hui $\&$ Becker 2006). Both of them have tail-like structures 
behind the moving pulsars. These trails could be associated 
with the shocked relativistic wind confined by the ram pressure 
of the ambient ISM. 

The XMM-Newton data of PSR B1957+20 have provided observational 
evidence for a strong dependence of the pulsar's X-ray emission 
on its binary orbital phase. It is the first time that a significant 
X-ray flux modulation from PSR B1957+20 is observed. 
The emission near to the radio eclipse is supposed to be beamed in a 
forward cone because the shocked fluid is accelerated by the pressure 
gradient as it flows around the eclipse region. Relativistic beaming 
would tend to give the maximum flux just before and after the pulsar 
eclipse which for PSR B1957+20 is at orbital phase = 0.25. Arons $\&$ 
Tavani (1993) predicted that the immediately downstream flow velocity 
of the shocked pulsar wind from the shock area along the line between 
the pulsar and the ablating companion is about c/3 and is even 
higher behind the relativistic shock when it passes around the 
companion. The tenuous relativistic plasma accelerates as it 
flows around the companion, possibly passing through a sonic 
transition to leave the binary system with the velocity larger 
than $c / \sqrt{3}$. Due to Doppler boosting, the probable 
post-shock velocities of the relativistic wind then suggest X-ray 
emission variation around the orbital phase by a factor between 
1.3 ($v= c/3$) and 2.2 ($v \cong c/\sqrt{3}$). This numbers were
estimated without considering the effect of absorption and/or 
scattering within the binary. On the contrary, the X-ray emission 
at the eclipse may be reduced because of the obscuration of the 
shock by the companion.

However, given the limited photon statistics of the XMM-Newton data 
it is not possible to investigate any binary-phase resolved imaging 
and spectral variation as a function of orbit phase or to determine 
the exact geometry of the peak emission. Such analysis would allow 
us to investigate with higher accuracy than currently possible whether 
the X-ray emission from PSR B1957+20 is present during all orbit angles 
or virtually only near to the radio eclipse while diffuse X-ray emission 
from the PWN is present at any orbital angle. Indeed, the later is 
suggested by the current data and a confirmation would have a severe 
impact on our understanding of the pulsar's X-ray emission properties. 

As the present XMM observation covers barely one binary orbit, we stress
that it can not be fully excluded that the increase in photon flux near 
the orbital angle of the radio eclipse is due to a single ``burst like" 
event or that the peak flux emission varies from orbit to orbit or on 
longer time scales. Clearly, a repeated coverage of the binary orbit 
in a longer XMM-Newton observation and a comparison with the 2004 data 
would answer this question immediately. This, in addition, 
would provide us not only a better photon statistic but would also allow 
us to determine the emission geometry with a much higher accuracy than 
currently possible.

\begin{acknowledgements}
This work made use of the XMM-Newton data archive. The first author 
acknowledges the recipe of funding provided by the Max-Planck Society in 
the frame of the International Max-Planck Research School (IMPRS). 
\end{acknowledgements}


\begin{thebibliography}{}

   \bibitem[1982]{Alpar} Alpar, M. A., Cheng, A. F., Ruderman, M. A., Shaham, J., 1982, Nature, 300, 728
   
   \bibitem[1993]{Arons} Arons, J. \& Tavani, M., 1993, ApJ, 403, 249
  
   \bibitem[2002]{Becker} Becker, W. \& Aschenbach, B., 2002, in Proceedings of the
   WE-Heraeus Seminar on Neutron Stars, Pulsars and Supernova remnants,
   eds.~W. Becker, H. Lesch \& J. Tr\"umper, MPE-Report 278, 64,
   (arxiv:astro-ph/0208466)
   
   \bibitem[1976]{Blandford} Blandford, R., Teukolsky, S. A., 1976, ApJ, 205, 580
   
   \bibitem[1983]{Buccheri} Buccheri, R., Bennett, K., Bignami, G. F., Bloemen, J. B. G. M., 
   Boriakoff, V., Caraveo, P. A., Hermsen, W., Kanbach, G., Manchester, R. N., Masnou, J. L., 
   Mayer-Hasselwander, H. A., Ozel, M. E., Paul, J. A., Sacco, B., Scarsi, L., Strong, A. W., 
   1983, A\&A, 128, 245 

   \bibitem[2006]{Bogdanov} Bogdanov, S., Grindlay, J. E., Heinke, C. O., Camilo, F., Freire, P. C. C., 
   Becker, W., 2006, ApJ, 646, 1104

   \bibitem[1988]{Fruchter} Fruchter, A. S., Gunn, J. E., Lauer, T. R., Dressler, A., 1988, Nature, 334, 686

   \bibitem[1988]{Fruchter1} Fruchter, A. S., Stinebring, D. R., Taylor, J. H. 1988, Nature, 333, 237

   \bibitem[2006]{Gaensler} Gaensler, B. M. \& Slane, P.O., 2006, Ann. Rev. Astron. Astrophys., 44, 17 

   \bibitem[2006]{Hui} Hui, C. Y. \& Becker, W., 2006, A$\&$A, 448, L13

   \bibitem[2006]{Kargaltsev} Kargaltsev, O. Y.\& Pavlov, G. G., in preparation (2006)

   \bibitem[2006]{Kaspi} Kaspi, V. M., Roberts, M. S. E., Harding, A. K., In: Compact Stellar X-ray Sources, 
   ed. W. H. G. Lewin \& M. van der Klis, Cambridge University Press, p. 279, 2006

   \bibitem[2001]{Krongold} Krongold, Y., Dultzin-Hacyan, D., Marziani, P., 2001, AJ, 121, 702

   \bibitem[1988]{Kulkarni} Kulkarni, S. R., Hester, J. J., 1988, Nature, 335, 801
   
   \bibitem[1992]{Kulkarni1} Kulkarni, S. R., Phinney, E. S., Evans, C. R., \& Hasinger, G., 1992, Nature, 359, 300
   
   \bibitem[2005]{Manchester} Manchester, R. N., Hobbs, G. B., Teoh, A., Hobbs, M., 2005, AJ, 129, 1993

   \bibitem[2003]{Monet} Monet, D. G., et al., 2003, AJ, 125, 984

   \bibitem[2003]{Stappers} Stappers, B. W., Gaensler, B. M., Kaspi, V. M., van der Klis, M., 
   \& Lewin, W. H. G., 2003, Science, 299, 1372
   
   \bibitem[1993]{Taylor} Taylor, J.H. \& Cordes, J.M., 1993, ApJ, 411, 674

   \bibitem[1988]{Paradijs} van Paradijs, J., Allington-Smith, J., Callanan, P., Hassall, B. J. M., 
   Charles, P. A, 1988, Nature, 334, 684


\end{thebibliography}
\end{document}